\documentclass[10pt]{article}

\usepackage[margin=18mm]{geometry}
\usepackage{authblk}
\usepackage{microtype}
\usepackage{amsmath,amssymb,bm}
\usepackage{booktabs,tabularx,array,multirow}
\usepackage{graphicx}
\usepackage{xcolor}
\usepackage{tikz}
\usetikzlibrary{arrows.meta,positioning}
\usepackage[numbers,sort&compress]{natbib}
\usepackage[hidelinks]{hyperref}
\usepackage[nameinlink,noabbrev]{cleveref}
\usepackage[font=small,labelfont=bf]{caption}
\usepackage{siunitx}
\usepackage[utf8]{inputenc}
\usepackage[T1]{fontenc}

\definecolor{cztblue}{HTML}{1769AA}
\definecolor{pipsorange}{HTML}{D95F02}
\definecolor{backgreen}{HTML}{2B8C4B}
\definecolor{warningred}{HTML}{B2182B}
\definecolor{softgray}{HTML}{F2F4F6}

\newcolumntype{Y}{>{\raggedright\arraybackslash}X}
\newcolumntype{C}{>{\centering\arraybackslash}X}
\newcommand{\sensunit}{\mathrm{cps}/(\mu\mathrm{Sv}/\mathrm{h})}
\newcommand{\Sone}{S1}
\newcommand{\Stwo}{S2}
\newcommand{\vect}[1]{\bm{#1}}

\title{\textbf{Physics-Constrained Co-Design and Data-Driven Layer-Resolved Classification of a Hybrid CZT/PIPS Detector for Mixed Radiation Fields}}
\author[1]{Renlong~Jie\thanks{Email: jierenlong@nwpu.edu.cn}}
\author[2]{Fan~Yang}
\author[2]{Shouzhi~Xi}
\author[2]{Sanqi~Tang}
\author[1,2]{Wanqi~Jie}
\affil[1]{Northwestern Polytechnical University, Xi'an 710072, China}
\affil[2]{Shaanxi Imdetek Co., Ltd., Xianyang 712000, China}
\date{}

\begin{document}
\maketitle

\begin{abstract}
Compact mixed-radiation instruments must preserve a low-mass charged-particle entrance while providing enough high-\(Z\) depth for photon sensitivity. We first compare two detector heads within a \(40\times20\times10\ \mathrm{mm^3}\) design budget. \Sone\ places bare CdZnTe (CZT) and passivated implanted planar silicon (PIPS) branches side by side and estimates three rates. \Stwo\ adds \(0.50\ \mathrm{mm}\) of CZT behind PIPS to estimate X/gamma, low-beta, high-beta, and alpha rates. A hard-constraint search combines photon attenuation and deposition, Hecht charge collection, charged-particle energy loss, solid angle, resolution budgeting, timing, and response-matrix conditioning. Feasible screening points exist inside the initial envelope, but a fresh transport/electronics assessment gives only 25.04--25.07 \(\sensunit\) under the robust \(H^*(10)\) convention, and every conservative electronics draw exceeds 2.5\% FWHM. We therefore retain the more informative \Stwo\ observation structure, relax only the 10 mm package-depth constraint, divide the bare CZT into independently biased layers, and add a low-noise sum channel for spectroscopy plus layer-resolved gradient-boosted classification for mixed-field analysis. The extension campaign contains 2.40 million Geant4 11.4.1 histories over three depths, five transport seeds, and 14 particle/energy cases. The final \(40\times20\times12.570\ \mathrm{mm^3}\) head uses 7.870 mm of bare CZT in five layers at 180 V per layer. Under the application-scenario U95 electronics profile, its live-time-corrected 662 keV sensitivity is 32.804 \(\sensunit\), its independent-sum resolution is 2.303\% FWHM at the 95th percentile, and the propagated beta/alpha absolute-efficiency lower bounds are 35.679\%/39.832\%. Additional endpoint transport covers 20 keV--3 MeV photons, 3--7 MeV alpha particles, and 155 keV--3.5 MeV beta spectra. The extended design therefore passes the original detector-performance criteria in the U95 model. 
\end{abstract}


\section{Introduction}

Room-temperature CZT combines high photon stopping power with direct charge conversion, but carrier trapping, electrode geometry, and crystal nonuniformity couple absorber thickness to spectroscopic quality \citep{schlesinger2001czt,owens2004compound,delsordo2009czt}. Thin-window silicon provides the complementary charged-particle channel. The design problem is both spatial and physical: the same head must preserve a low-mass entrance for alpha and beta particles, allocate enough high-\(Z\) depth for photons, and expose enough independent observables to estimate overlapping field components.

The initial detector-head budget is \(4\ \mathrm{cm}\times2\ \mathrm{cm}\times1\ \mathrm{cm}\). The project brief specifies X/gamma coverage from 20 keV to 3 MeV, 662 keV sensitivity of at least 30 \(\sensunit\), 662 keV energy resolution no worse than 2.5\% FWHM, alpha coverage from 3 to 7 MeV with at least 25\% absolute efficiency, and beta coverage from 155 keV to 3.5 MeV with at least 35\% absolute efficiency. The photon-sensitivity denominator is interpreted as the radiation-protection quantity \(H^*(10)\); an air-kerma-equivalent convention is retained only to reconcile the earlier optimization.

The study proceeds in two design phases. The first keeps the complete \(40\times20\times10\ \mathrm{mm^3}\) budget fixed. It compares a parallel three-class head with a parallel-plus-stacked four-class head through physics-constrained search, particle transport, electronics perturbation, and finite-count response-matrix closure. That phase identifies feasible screening points but does not close the robust \(H^*(10)\) sensitivity and conservative resolution requirements simultaneously. The second phase retains the 40 by 20 mm face and the \Stwo\ sensing principle, relaxes only package depth, and co-optimizes total CZT thickness, number of independently biased layers, and continuous-energy signal processing. The resulting design is evaluated with a multi-seed transport campaign and a U95 application electronics profile. The project verdict uses the original detector-performance criteria; response-matrix conditioning and mixture-recovery error remain supporting algorithm metrics.

\section{Detector context and related work}

CZT spectrometers often use electrode design or depth correction to reduce the influence of poorly transported holes. Coplanar single-polarity sensing and small-pixel weighting potentials are established examples \citep{luke1994single,barrett1995charge}, while detector reviews emphasize that mobility--lifetime products and depth-dependent charge loss remain central design variables \citep{eisen1999cdte,schlesinger2001czt}. These observations motivate the layered, independently biased CZT stack and the explicit charge-acceptance term used here; absorber thickness is not treated as an unconditional gain in usable sensitivity.

For silicon charged-particle detectors, entrance dead layers shape energy loss, while source--detector geometry sets the accepted solid angle and therefore the counting efficiency. Measurements and Geant4 studies of PIPS and position-sensitive silicon devices show why these quantities must be characterized rather than absorbed into a generic intrinsic-efficiency constant \citep{diazfrances2017pips,phong2018pips,manfredi2018deadlayer}. Stacked \(\Delta E\)--\(E\) telescopes provide the broader physical precedent for using correlated energy deposition in two layers to separate charged-particle populations \citep{tassangot2002deltae,topkar2011deltae}. The present PIPS/back-CZT branch applies that principle to two beta spectra, then combines its gates with a separately illuminated bare-CZT photon branch.

The beta source model also matters. Allowed-spectrum formulae include Coulomb, finite-size, radiative, atomic, and recoil corrections \citep{hayen2018beta}; forbidden transitions can depart from the simplest allowed shape \citep{mougeot2015beta}. We therefore describe the simulated inputs by their endpoint spectra rather than treating an endpoint energy as a monoenergetic electron. The resulting response matrix is recovered with a nonnegative Poisson likelihood, in the same family as established emission-reconstruction methods \citep{shepp1982mle}.

\section{Architectures and performance requirements}

\subsection{Two detector-head structures}

Both structures allocate a \(17\times18\ \mathrm{mm^2}\) charged-particle aperture and a \(20\times18\ \mathrm{mm^2}\) photon aperture, separated by a 1 mm grounded rib and surrounded by 1 mm package margins. The photon branch contains three independently biased CZT layers with total active thickness 5.5 mm. The charged-particle branch contains a 2 \(\mu\)m low-mass window and 350 \(\mu\)m of silicon.

In \Sone, the PIPS and bare-CZT branches are parallel. Exclusive gates produce three recovered rate components: X/gamma, beta, and alpha. In \Stwo, a 0.50 mm CZT layer is placed behind the PIPS. A low PIPS pulse without backing coincidence forms the low-beta gate, while a low/intermediate PIPS pulse with backing coincidence forms the high-beta gate. High PIPS energy with backing anticoincidence forms the alpha gate. Bare-CZT and stack-only photon gates complete a five-row response matrix for four rate components. The architecture does not identify every particle without error. It estimates class count rates from a calibrated mixed response.

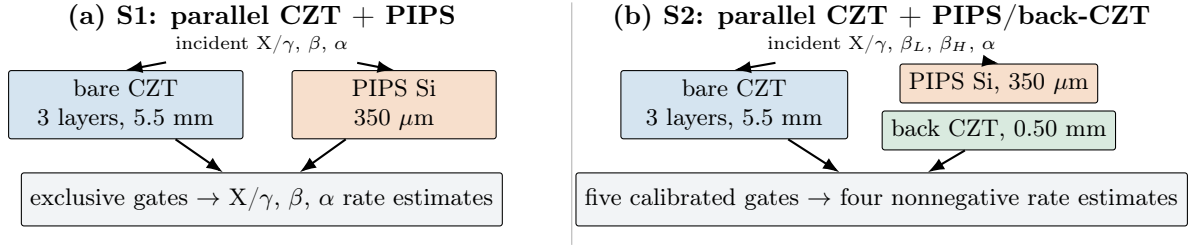
\begin{figure}[t]
\centering
\begin{tikzpicture}[
  font=\small,
  box/.style={draw,rounded corners=1pt,minimum height=7mm,align=center},
  flow/.style={-{Latex[length=2.2mm]},thick},
  note/.style={font=\scriptsize,align=center}
]
  \node[font=\bfseries] at (0,2.35) {(a) S1: parallel CZT + PIPS};
  \node[box,fill=cztblue!18,minimum width=31mm] (s1czt) at (-1.8,1.20) {bare CZT\\3 layers, 5.5 mm};
  \node[box,fill=pipsorange!20,minimum width=27mm] (s1si) at (1.75,1.20) {PIPS Si\\350 $\mu$m};
  \node[note] (s1rad) at (0,2.00) {incident X/$\gamma$, $\beta$, $\alpha$};
  \draw[flow] (s1rad) -- (s1czt.north);
  \draw[flow] (s1rad) -- (s1si.north);
  \node[box,fill=softgray,minimum width=55mm] (s1out) at (0,-0.05) {exclusive gates $\rightarrow$ X/$\gamma$, $\beta$, $\alpha$ rate estimates};
  \draw[flow] (s1czt) -- (s1out);
  \draw[flow] (s1si) -- (s1out);

  \draw[gray!60] (4.1,-0.65) -- (4.1,2.55);

  \node[font=\bfseries] at (8.2,2.35) {(b) S2: parallel CZT + PIPS/back-CZT};
  \node[box,fill=cztblue!18,minimum width=30mm] (s2czt) at (6.25,1.20) {bare CZT\\3 layers, 5.5 mm};
  \node[box,fill=pipsorange!20,minimum width=25mm,minimum height=5mm] (s2si) at (9.75,1.48) {PIPS Si, 350 $\mu$m};
  \node[box,fill=backgreen!18,minimum width=25mm,minimum height=5mm] (s2back) at (9.75,0.86) {back CZT, 0.50 mm};
  \node[note] (s2rad) at (8.2,2.00) {incident X/$\gamma$, $\beta_L$, $\beta_H$, $\alpha$};
  \draw[flow] (s2rad) -- (s2czt.north);
  \draw[flow] (s2rad) -- (s2si.north);
  \node[box,fill=softgray,minimum width=66mm] (s2out) at (8.2,-0.05) {five calibrated gates $\rightarrow$ four nonnegative rate estimates};
  \draw[flow] (s2czt) -- (s2out);
  \draw[flow] (s2back) -- (s2out);
\end{tikzpicture}
\caption{Detector concepts. The two branches are side by side on the incident face. The backing CZT in \Stwo\ adds an observation dimension for statistical beta-band separation; it is not a perfect event-by-event particle label.}
\label{fig:architectures}
\end{figure}

\subsection{Performance constraints and evidence mapping}

\Cref{tab:requirements} lists the frozen requirements used for screening. Feasibility is defined by their intersection; a high value on one metric cannot compensate for a failed constraint on another.

\begin{table}[t]
\centering
\caption{Design requirements and the evidence needed to close them.}
\label{tab:requirements}
\small
\begin{tabularx}{\textwidth}{p{31mm}p{35mm}Y}
\toprule
Requirement & Screening constraint & Evidence status in this study \\
\midrule
Detector-head envelope & \(40\times20\times10\ \mathrm{mm^3}\) & Dimensioned model; electronics outside the head are excluded \\
X/gamma range & 20 keV--3 MeV & Endpoint constraints plus six discrete Geant4 photon energies \\
Sensitivity at 662 keV & \(S_{662}\geq30\ \sensunit\) & Effective-count model evaluated for both \(K_a\) and \(H^*(10)\) conventions \\
Energy resolution at 662 keV & FWHM \(\leq2.5\%\) & Inverse budget and electronics Monte Carlo; hardware acceptance specification \\
Alpha range and efficiency & 3--7 MeV; \(\eta_\alpha\geq25\%\) & Range/solid-angle model; Geant4 at 3 and 5.486 MeV \\
Beta range and efficiency & 0.155--3.5 MeV; \(\eta_\beta\geq35\%\) & Range/solid-angle model; Geant4 endpoints at 0.546 and 2.280 MeV \\
Four-class mixed-field output & X/gamma, \(\beta_L\), \(\beta_H\), alpha & Required only for \Stwo; evaluated as rate unmixing, not perfect classification \\
\bottomrule
\end{tabularx}
\end{table}

\section{Physics-constrained design model}

\subsection{Photon effective-count sensitivity}

The screening observable is the accepted live-count sensitivity
\begin{equation}
S_{\mathrm{eff}}(E,\vect{x})=
k_{\Phi}(E)A_XT_w(E)P_{\mathrm{int}}(E,\vect{x})
f_{\mathrm{dep}}(E,\vect{x})P_{\mathrm{cc}}(E,\vect{x})f_{\mathrm{live}},
\label{eq:sensitivity}
\end{equation}
where \(k_\Phi\) converts the stated dose-rate quantity to incident fluence rate, \(A_X\) is the photon aperture, \(T_w\) is entrance-window transmission, \(P_{\mathrm{int}}=1-\exp[-\mu(E)d]\) is the interaction probability, \(f_{\mathrm{dep}}\) is the finite-aperture nonzero-deposition fraction, \(P_{\mathrm{cc}}\) is fixed-gain charge-pulse acceptance, and \(f_{\mathrm{live}}\) is the live fraction. Mass-attenuation values are locally interpolated on a log--log scale from an embedded table based on NIST XCOM, then converted to linear coefficients with CZT and Al densities of 5.78 and 2.70~\(\mathrm{g\,cm^{-3}}\) \citep{nistxcom}. The finite-aperture factor is anchored to earlier energy-deposition calculations and bounded to \([0.75,0.98]\), which prevents every interaction from being counted as a usable pulse.

The dose-to-fluence term is fixed from
\begin{equation}
\dot K_a=\dot\Phi E\left(\frac{\mu_{\mathrm{tr}}}{\rho}\right)_{\!\mathrm{air}}
\simeq\dot\Phi E\left(\frac{\mu_{\mathrm{en}}}{\rho}\right)_{\!\mathrm{air}}.
\label{eq:kerma}
\end{equation}
At 662 keV, interpolation of the NIST dry-air mass energy-absorption coefficients gives \(k_{\Phi,K}=89.4\ \mathrm{photons\,cm^{-2}\,s^{-1}}\) per \(\mu\mathrm{Gy\,h^{-1}}\) \citep{hubbell1995xaamdi}. This numerical conversion uses collision kerma as the approximation to air kerma; their difference is the small radiative-loss fraction in air at this energy. The robust screen applies a 0.95 fluence factor, giving 84.9. Radiation-protection calibration instead uses an operational quantity. The 2000 IAEA calibration report tabulates \(H^*(10)/K_a=1.20\ \mathrm{Sv/Gy}\) for S-Cs, consistent with the neighboring monoenergetic coefficients in ICRP Publication 74 \citep{iaea2000calibration,icrp1996coefficients}. ISO 4037:2019 changes the S-Cs air-kerma-to-operational-quantity coefficient to 1.21 and supplies the current area-dosemeter calibration framework \citep{iso4037_3_2019,hupe2021iso4037}. We therefore use \(k_{\Phi,H^*}=73.9\) centrally and 70.2 after the same fluence factor.

\begin{table}[t]
\centering
\caption{Effect of dose convention on the selected screening-point sensitivity. The robust column also includes the model's attenuation, deposition, charge-acceptance, and live-time reductions.}
\label{tab:doseconvention}
\small
\begin{tabular}{llrrrr}
\toprule
Dose convention & Fluence factor, central/robust & \multicolumn{2}{c}{\Sone} & \multicolumn{2}{c}{\Stwo} \\
 & \(\mathrm{cm^{-2}s^{-1}}\) per unit dose rate & nominal & robust & nominal & robust \\
\midrule
Air kerma, \(K_a\) & 89.4 / 84.9 per \(\mu\mathrm{Gy\,h^{-1}}\) & 41.64 & 30.33 & 41.60 & 30.30 \\
Ambient dose, \(H^*(10)\) & 73.9 / 70.2 per \(\mu\mathrm{Sv\,h^{-1}}\) & 34.41 & 25.07 & 34.38 & 25.04 \\
\bottomrule
\end{tabular}
\end{table}

The robust calculation also uses a 0.95 attenuation scale, a 0.97 deposition multiplier, and \(f_{\mathrm{live}}=0.92\). These factors define the screening envelope. The Geant4 results are reported as a separate event-transport calculation rather than being used to retune \(f_{\mathrm{dep}}\).

\subsection{Charge collection and the resolution budget}

For layer thickness \(d_L\), bias \(V_L\), and interaction depth \(z\), we use the planar Hecht relation \citep{hecht1932transport}
\begin{align}
\eta_H(z) &= \frac{\lambda_e}{d_L}\left[1-\exp\left(-\frac{d_L-z}{\lambda_e}\right)\right]
+\frac{\lambda_h}{d_L}\left[1-\exp\left(-\frac{z}{\lambda_h}\right)\right],\\
\lambda_{e,h} &= \frac{(\mu\tau)_{e,h}V_L}{d_L}.
\label{eq:hecht}
\end{align}
This planar approximation follows standard charge-induction treatments \citep{he2001shockley}. Nominal electron/hole mobility-lifetime products are \(10^{-3}/10^{-5}\ \mathrm{cm^2\,V^{-1}}\), with screening floors of \(3\times10^{-4}/3\times10^{-6}\ \mathrm{cm^2\,V^{-1}}\). A depth-weighted threshold is calibrated to the 8 mm fixed-gain acceptance anchor used in the earlier model. The selected 5.5 mm stack consequently specifies the material and correction performance that a prototype must reproduce. Weighting-potential effects motivate the later hardware-calibration step \citep{luke1994single,barrett1995charge}; that step also addresses charge sharing, field nonuniformity, polarization, and interlayer scatter outside the planar screen.

The 662 keV resolution budget is
\begin{equation}
R_{\mathrm{FWHM}}(E)=\frac{2.355}{E}
\sqrt{FwE+(w\,\mathrm{ENC})^2+\sigma_{\mathrm{cc,res}}^2+\sigma_{\mathrm{cal}}^2},
\label{eq:resolution}
\end{equation}
where \(F=0.10\), \(w=4.64\ \mathrm{eV}\), ENC is the equivalent noise charge, and the last two terms represent residual charge-collection and calibration dispersion. The optimizer solves \Cref{eq:resolution} for the maximum admissible \(\sigma_{\mathrm{cc,res}}\). For \Sone, the non-charge-collection terms contribute 0.809\% FWHM, the remaining standard-deviation allowance is 1.004\% of energy, and the raw accepted-event Hecht variation must be reduced by a factor of 0.0509. This factor is a quantitative requirement on depth/pixel correction and calibration.

\subsection{Charged particles and geometric efficiency}

For a centered point source at distance \(r\) from a rectangular aperture with half-widths \(a\) and \(b\), the accepted fraction of an isotropic field is
\begin{equation}
\eta_{\mathrm{geo}}=\frac{1}{\pi}\tan^{-1}\left(\frac{ab}{r\sqrt{r^2+a^2+b^2}}\right).
\label{eq:solidangle}
\end{equation}
Absolute charged-particle efficiency is modeled as \(\eta_{\mathrm{abs}}=\eta_{\mathrm{geo}}\eta_{\mathrm{intrinsic}}\). Entrance-window and silicon losses use stopping-power and continuous-slowing-down range relations informed by NIST STAR tables \citep{niststar}. The alpha constraint requires at least 1.2 MeV after the low-mass window for a 3 MeV incident alpha. The beta constraint requires at least 40 keV deposited in silicon for a 155 keV electron. This construction keeps point-source geometry separate from the conditional intrinsic probabilities later reported for face-illuminated Geant4 events.

\subsection{Response matrix and Poisson rate unmixing}

Let \(\vect{R}\in\mathbb{R}_+^{G\times K}\) be the exclusive-gate response matrix, \(\vect{\Phi}\in\mathbb{R}_+^K\) the unknown class counts, and \(\vect{y}\) the observed gate counts. The virtual prototype uses
\begin{equation}
y_g\sim\mathrm{Poisson}([\vect{R}\vect{\Phi}]_g),\qquad
\widehat{\vect{\Phi}}=\arg\max_{\vect{\Phi}\geq0}
\sum_g\left[y_g\log([\vect{R}\vect{\Phi}]_g)-[\vect{R}\vect{\Phi}]_g\right].
\label{eq:poisson}
\end{equation}
The nonnegative estimate is computed with a multiplicative maximum-likelihood update related to established Poisson reconstruction methods \citep{shepp1982mle}. Background is set to zero in the present trials. The response-matrix rank and condition number are treated as design constraints because subtraction alone can amplify calibration and counting noise.

\section{Discrete search and frozen comparison design}

\subsection{Search procedure}

The \Sone\ search evaluates 8,640 combinations spanning 2.0--3.0 \(\mu\)m PIPS windows, 350--400 \(\mu\)m Si, 4.0--6.0 mm total CZT, one to three CZT layers, 200--600 V per layer, 0.20--0.25 mm Al photon windows, 1.0--1.2 mm ribs, 16--22 mm photon widths, and 1.0--2.0 mm pixel pitch. Of these grid points, 192 satisfy every hard constraint; none uses a single CZT layer. The reported design is therefore a selected feasible grid point, not a continuous or global optimum.

The \Stwo\ search freezes the selected bare-CZT baseline and evaluates 8,640 combinations of Si thickness, backing-CZT thickness, PIPS and backing thresholds, coincidence half-width, and bare-CZT window. It requires full column rank, condition number no larger than 20, bounded equal-flux gate impurity, maximum Fisher relative standard deviation no larger than 8\% at \(10^4\) counts per class, accidental coincidence probability below 1\%, and all shared performance constraints. There are 4,320 passing grid points. A secondary cost and margin criterion selects the reported point from this feasible set.

\subsection{Selected screening points and common comparison geometry}

\Cref{tab:designs} separates the screening checkpoints from the geometry used in the V4 comparison. The \Stwo\ search selected a 0.25 mm bare-CZT Al window with a semi-analytic condition number of 6.81. The subsequent comparison uses 0.20 mm for both structures, which removes the window difference and gives each head an occupied depth of 8.95 mm. The nominal V4 response matrix is recalculated from event transport and electronics propagation; its \Stwo\ condition number of 9.50 is therefore distinct from the semi-analytic screening value.

\begin{table}[t]
\centering
\caption{Selected detector-head parameters. The last two columns give the common V4 comparison geometry.}
\label{tab:designs}
\small
\begin{tabularx}{\textwidth}{p{39mm}p{28mm}p{28mm}YY}
\toprule
Parameter & \Sone\ screening & \Stwo\ screening & \Sone\ V4 & \Stwo\ V4 \\
\midrule
Package & \multicolumn{4}{c}{\(40\times20\times10\ \mathrm{mm^3}\)} \\
PIPS window / Si & 2.0 \(\mu\)m / 350 \(\mu\)m & 2.0 \(\mu\)m / 350 \(\mu\)m & same & same \\
Bare CZT active area & \(20\times18\ \mathrm{mm^2}\) & \(20\times18\ \mathrm{mm^2}\) & same & same \\
Bare CZT stack & 3 \(\times\) 1.833 mm & 3 \(\times\) 1.833 mm & same & same \\
Bias per CZT layer & 300 V & 300 V & same & same \\
Bare-CZT Al window & 0.20 mm & 0.25 mm & 0.20 mm & 0.20 mm \\
Backing CZT & none & 0.50 mm & none & 0.50 mm \\
PIPS low / alpha threshold & 40 / 800 keV & 40 / 800 keV & 40 / 800 keV & 40 / 800 keV \\
Backing threshold / coincidence & not applicable & 75 keV / \(\pm300\) ns & not applicable & 75 keV / \(\pm300\) ns \\
Occupied head depth & 8.95 mm & 9.00 mm & 8.95 mm & 8.95 mm \\
Robust \(S_{662}\), \(K_a\) convention & 30.33 & 30.30 & --- & --- \\
Semi-analytic point-source \(\eta_\alpha/\eta_\beta\) & 39.12\% / 39.84\% & 39.57\% / 39.80\% & --- & --- \\
\bottomrule
\end{tabularx}
\end{table}

The screening points leave 1.05 and 1.00 mm of depth headroom; the common V4 geometry leaves 1.05 mm for either structure. This envelope covers the detector head and its internal layers. High-voltage supply, acquisition electronics, and external connectors are system-level items outside the modeled head.

\subsection{Depth-relaxed \Stwo\ candidates}

The fixed-envelope study identifies \Stwo\ as the more informative architecture, but its robust \(H^*(10)\) sensitivity remains below the project threshold. We therefore retain the 40 by 20 mm incident face and relax only package depth. The bare-CZT branch is extended, whereas the \(17\times18\ \mathrm{mm^2}\) PIPS aperture, 350 \(\mu\)m silicon, 2 \(\mu\)m entrance window, 0.50 mm backing CZT, and 1 mm separator remain fixed. The additional CZT is divided into independently biased layers. This keeps the layer drift distance and field within the range used by the charge-acceptance model instead of treating a thicker monolithic crystal as an unconditional sensitivity gain.

Three geometries are carried into the fresh validation campaign (\Cref{tab:extendeddesigns}). The 11.577 mm candidate is the first depth-grid point whose raw analytical sensitivity reaches 30 \(\sensunit\). The 12.570 mm candidate reaches the 33 \(\sensunit\) engineering target before live-time correction and is used as the margin design. The initial 10 mm envelope is a design-space constraint rather than one of the detector-performance values listed in the project completion table; the depth-relaxed candidates make that trade explicit.

\begin{table}[t]
\centering
\caption{Depth-relaxed \Stwo\ candidates. Only the bare-CZT branch changes.}
\label{tab:extendeddesigns}
\small
\begin{tabularx}{\textwidth}{YCCCCY}
\toprule
Candidate & Package (mm) & Bare CZT & Layers & Bias/layer & Design role \\
\midrule
D10 baseline & \(40\times20\times10.000\) & 5.500 mm & 3 & 300 V & Initial package limit \\
D11.577 formal & \(40\times20\times11.577\) & 6.977 mm & 4 & 225 V & First 30-\(\sensunit\) raw crossing \\
D12.570 margin & \(40\times20\times12.570\) & 7.870 mm & 5 & 180 V & U95 operational margin candidate \\
\bottomrule
\end{tabularx}
\end{table}

\section{Three-stage computational assessment}

\subsection{Level 1: event-level particle transport}

Event transport was performed with the Geant4 toolkit \citep{agostinelli2003geant4,allison2016geant4}. The frozen build reports release 11.4.1, uses the \texttt{geant4\_pybind} package, and combines FTFP\_BERT with \texttt{G4EmStandardPhysics\_option4} (EMZ). The geometry contains a polyimide entrance window, silicon, three CZT layers separated by 100 \(\mu\)m gaps, optional backing CZT, a 0.20 mm Al photon window, and a copper separator. CZT is represented at 5.78 \(\mathrm{g\,cm^{-3}}\) with mass fractions 0.433 Cd, 0.028 Zn, and 0.539 Te.

Primary particles are normally incident and uniformly sampled over the two active faces in proportion to face area; the center rib is excluded. Photon energies are 20, 59.5, 122, 662, 1332, and 3000 keV, with 4,000 histories per point and structure. Alpha energies are 3 and 5.486 MeV, with 4,000 histories each. Following the continuous beta-spectrum form rather than substituting monoenergetic endpoints \citep{hayen2018beta,mougeot2015beta}, the code samples 6,000 beta-minus histories for each of the 0.546 and 2.280 MeV endpoint distributions. The campaign contains 88,000 event rows across 20 structure/source cases. Deposited energy is recorded separately in the PIPS window, silicon, bare CZT, backing CZT, Al window, and separator. Trigger proportions are accompanied by Wilson 95\% intervals \citep{wilson1927interval}.

The full-face photon probability includes the bare and stack branches under the common illumination. Charged-particle PIPS probabilities are conditioned on histories incident on the PIPS face. They are intrinsic geometry/material responses and must not be compared directly with the point-source absolute efficiencies in \Cref{tab:designs}.

\subsection{Level 2: electronics uncertainty propagation}

Level 2 uses the 662 keV photon pool, the 5.486 MeV alpha pool, and the two beta pools. For \Sone, the beta pools are combined into one class; \Stwo\ retains low- and high-beta classes. Channel energy is perturbed according to
\begin{equation}
\sigma_E^2=FwE+\sigma_{\mathrm{ENC}}^2+(gE)^2+(cE)^2,
\label{eq:electronics}
\end{equation}
where \(g\) and \(c\) represent gain and charge-collection dispersion. Crosstalk is modeled as conservative charge mixing between adjacent channel pairs. Thresholds drift independently. True PIPS/back-CZT coincidences are accepted using a Gaussian timing-jitter model, while accidental coincidence probability is \(p_{\mathrm{acc}}=\min(1,2r\tau_c)\). The nonparalyzable live fraction is \(L=(1+r\tau_d)^{-1}\).

Three profiles span nominal, conservative, and stress conditions. Each structure/profile pair receives 30 electronics draws from a shared structure-specific transport pool, giving 180 response matrices. Nominal/conservative/stress bare-CZT ENC values are 2.5/5/10 keV and back-CZT values are 3/6/12 keV. Gain dispersions are 0.25/0.5/1.0\%, charge-collection dispersions are 0.45/0.85/1.25\%, modeled crosstalk fractions are 0.2/0.7/1.5\%, rates are 1/5/10 kcps, and dead times are 2/5/10 \(\mu\)s. The profiles are engineering envelopes used to expose transition points in resolution and matrix conditioning.

\subsection{Level 3: finite-count same-model closure}

For each structure/profile, the mean Level-2 response matrix is treated as the latent response. A calibration matrix is sampled from it with 20,000 multinomial events per source class, and test observations are sampled from the same latent matrix. This construction isolates finite calibration statistics, Poisson counting noise, and solver behavior under a correctly specified response; it does not introduce transport-model mismatch. Five labeled mixtures are evaluated: balanced, photon-dominant, low-beta-dominant, high-beta-dominant, and alpha-trace. After the two beta classes are combined, \Sone\ has four distinct truth vectors because the two beta-dominant cases coincide. Incident totals are \(10^3\), \(10^4\), and \(10^5\), with 200 trials per labeled combination. Across two structures, three profiles, five labels, and three count levels, the retained table contains 18,000 fits. Each fit uses at most 600 nonnegative multiplicative iterations with a relative \(\ell_1\) stopping tolerance of \(10^{-10}\).

For \Sone, the two beta components are collapsed before recovery, so \(K=3\). For \Stwo, \(K=4\). Error is reported as
\begin{equation}
\mathrm{RMSE}_{\mathrm{vec}}=\frac{100\%}{N}
\sqrt{\frac{1}{K}\sum_{k=1}^{K}(\widehat{\Phi}_k-\Phi_k)^2},
\label{eq:rmse}
\end{equation}
where \(N\) is the total incident count. This normalization avoids an unbounded percentage error for trace components, but component-wise percentage-point errors are also retained.

\subsection{Depth-relaxed U95 transport and signal processing}

The depth-relaxed campaign evaluates D10, D11.577, and D12.570 with five Geant4 seeds (202608201--202608205). Each geometry/seed pair covers seven photon energies from 20 keV to 3 MeV, alpha particles at 3, 5.486, and 7 MeV, and beta spectra with 155 keV, 0.546 MeV, 2.280 MeV, and 3.5 MeV endpoints. Photon and alpha cases contain 10,000 histories; beta cases contain 15,000. The campaign therefore contains 2.40 million event rows in 210 completed jobs. Bare-CZT deposition is stored both as a total and by layer; their largest event-wise difference is \(1.33\times10^{-15}\) MeV.

The seed roles are fixed before the final assessment: 202608201 trains the classifiers, 202608202 and 202608203 form the calibration pool, 202608204 is reserved for development-stage threshold-semantics checks, and 202608205 is the locked confirmation transport set. The U95 application profile uses Si/bare-CZT/back-CZT ENC values of 6.4/3.5/4.0 keV, relative gain and charge-collection dispersions of 0.004 and 0.0065, 0.4\% modeled crosstalk, 2 keV threshold drift, 75 ns timing jitter, 3 \(\mu\)s dead time, and a 2 kcps event rate. Twenty electronics realizations are evaluated per geometry.

Three signal-processing paths are compared. Fixed gates use the same physical thresholds as the first campaign. Aggregate gradient-boosted decision trees use continuous PIPS, bare-sum, and backing-CZT energies. The layered model adds individual layer energies, their fractions, an energy-weighted layer centroid, and layer multiplicity. A separate low-noise sum channel supplies the 662 keV spectrum; the layer channels are retained for classification. This division is necessary because summing independently noisy layer channels does not reproduce the sum-channel resolution.

The extension-stage project gate contains only the detector-performance criteria in the project completion table: live-time-corrected 662 keV sensitivity of at least 30 \(\sensunit\), 662 keV FWHM no greater than 2.5\%, alpha and beta absolute-efficiency lower bounds of at least 25\% and 35\%, and nonzero response at the specified energy-range endpoints. The 33 \(\sensunit\) target, response-matrix condition number, classifier F1, and mixture-recovery errors are engineering and algorithm diagnostics. In particular, the 5\% total-bias line used during algorithm development is not a project acceptance criterion.

\section{Results}

\subsection{Level-1 transport response}

\begin{figure}[t]
\centering
\includegraphics[width=\textwidth]{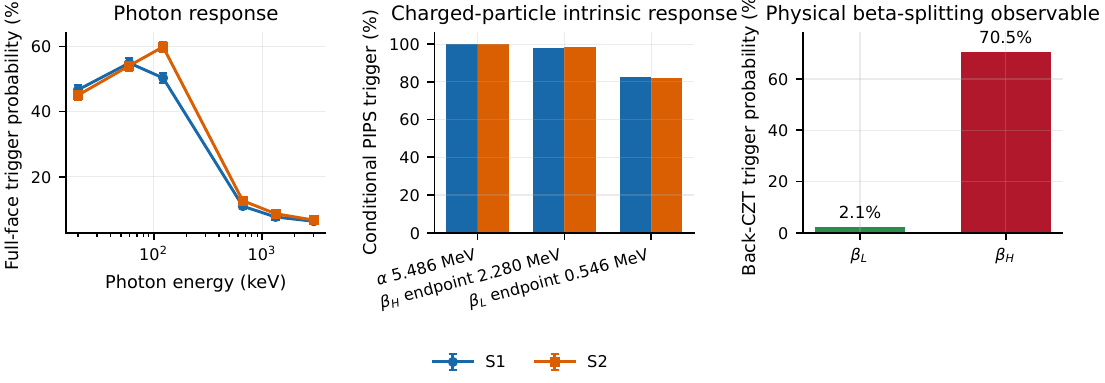}
\caption{Level-1 Geant4 transport. Photon error bars are Wilson 95\% intervals. Charged-particle PIPS trigger probabilities are conditional on incidence on the PIPS face. The right panel isolates the additional backing-CZT beta observable in \Stwo.}
\label{fig:transport}
\end{figure}

\Cref{fig:transport,tab:level1} show the event-level transport results. The nonmonotonic low-energy photon response is physically consistent with the division between a thick bare-CZT branch and a thin stack branch. The 0.50 mm backing CZT increases the 122 keV full-face trigger from 50.33\% to 59.88\%, while the improvement at 662 keV is 1.68 percentage points. At 3 MeV, both heads retain a nonzero response near 6--7\%.

\begin{table}[t]
\centering
\caption{Selected Level-1 trigger probabilities. Parentheses contain Wilson 95\% intervals.}
\label{tab:level1}
\small
\begin{tabularx}{\textwidth}{YCC}
\toprule
Metric & \Sone & \Stwo \\
\midrule
20 keV photon, full face & 46.68\% (45.13--48.22) & 45.03\% (43.49--46.57) \\
122 keV photon, full face & 50.33\% (48.78--51.87) & 59.88\% (58.35--61.38) \\
662 keV photon, full face & 11.00\% (10.07--12.01) & 12.68\% (11.68--13.74) \\
3 MeV photon, full face & 6.35\% (5.64--7.15) & 6.73\% (5.99--7.54) \\
0.546 MeV endpoint beta, PIPS conditional & 82.60\% (81.14--83.97) & 82.23\% (80.75--83.61) \\
2.280 MeV endpoint beta, PIPS conditional & 97.76\% (97.14--98.25) & 98.54\% (98.02--98.93) \\
5.486 MeV alpha, PIPS conditional & 100\% (99.79--100) & 100\% (99.79--100) \\
Backing CZT, low/high beta conditional & not applicable & 2.12\% / 70.47\% \\
\bottomrule
\end{tabularx}
\end{table}

The backing-CZT beta contrast is the key Level-1 evidence for four-class recovery: high-endpoint beta histories reach the backing sensor far more often than low-endpoint histories. Alpha particles do not trigger the backing sensor in the sampled normal-incidence geometry. The additional observable arises from energy-dependent transmission through PIPS, closely analogous to the stopping/range principle used in \(\Delta E\)--\(E\) telescopes; conventional telescopes additionally correlate upstream energy loss with downstream residual energy \citep{tassangot2002deltae,topkar2011deltae}.

The conditional PIPS probabilities can also update the point-source beta estimate. Multiplying the \(2\ \mathrm{mm}\) centered-source geometric fraction, 39.12\%, by the mean low/high-beta PIPS trigger probability gives 35.28\% for \Sone\ and 35.36\% for \Stwo\ under an equal-emission \(^{90}\)Sr/\(^{90}\)Y proxy. Propagating the endpoint-wise Wilson bounds gives 34.87--35.64\% and 34.97--35.70\%, respectively. The point estimates sit just above the 35\% requirement, while their lower bounds sit just below it. By contrast, the corresponding alpha estimate remains about 39.1\%, with a propagated lower bound near 39.0\%, comfortably above the 25\% requirement. These values apply to the modeled centered-source geometry.

\subsection{Level-2 electronics robustness}

\begin{figure}[t]
\centering
\includegraphics[width=\textwidth]{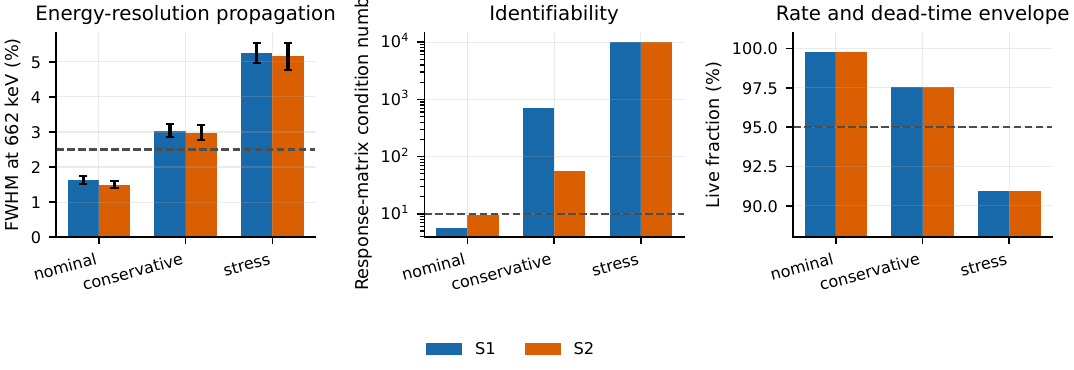}
\caption{Level-2 electronics Monte Carlo. Error bars show the standard deviation across 30 replicates for FWHM. The dashed guides mark the 2.5\% resolution budget, a condition-number reference of 10, and the 95\% live-fraction gate, respectively. Infinite or extreme stress-profile condition numbers are plotted at \(10^4\) for visibility.}
\label{fig:electronics}
\end{figure}

The nominal mean profiles satisfy the 2.5\% resolution budget and the condition-number gate of 10 for both structures (\Cref{fig:electronics,tab:level2}). One of the 30 nominal \Sone\ draws reaches 10.04; all nominal \Stwo\ draws remain below 10. Every conservative draw exceeds 2.5\% FWHM for both structures, and the mean condition number rises to 713 for \Sone\ and 55 for \Stwo. The stress results are strongly skewed: three of 30 \Sone\ matrices lose rank, whereas the mean matrix remains full rank with condition number 1287.5. For \Stwo, the stress median and 95th percentile are 124.6 and 591.1; the arithmetic mean of \(2.52\times10^{15}\) is driven by one \(7.57\times10^{16}\) outlier.

\begin{table}[t]
\centering
\caption{Electronics Monte Carlo summary. FWHM is mean \(\pm\) sample standard deviation. Condition numbers are means for nominal/conservative profiles and distribution summaries for stress profiles.}
\label{tab:level2}
\small
\begin{tabular}{llccc}
\toprule
Structure & Profile & 662 keV FWHM & Condition number & Live fraction \\
\midrule
\Sone & nominal & \(1.626\pm0.109\%\) & 5.69 & 99.80\% \\
\Sone & conservative & \(3.037\pm0.186\%\) & 713 & 97.56\% \\
\Sone & stress & \(5.257\pm0.291\%\) & 3/30 rank loss; finite median 2235 & 90.91\% \\
\Stwo & nominal & \(1.500\pm0.098\%\) & 9.50 & 99.80\% \\
\Stwo & conservative & \(2.976\pm0.209\%\) & 55.0 & 97.56\% \\
\Stwo & stress & \(5.157\pm0.384\%\) & median 124.6; p95 591.1 & 90.91\% \\
\bottomrule
\end{tabular}
\end{table}

The \Sone\ degradation is consistent with the modeled leakage of high-amplitude alpha signals into the low-threshold bare-CZT photon gate. Its alpha-to-bare-photon response rises from 0.0255 nominally to 0.447 under the conservative profile. This pattern makes analog isolation, controlled grounding, and a digital veto direct design priorities for \Sone. \Stwo\ retains an additional row, yet its conservative conditioning still calls for periodic single-source or pulser calibration and an online conditioning monitor. A crosstalk ablation on measured waveforms is the appropriate test of causality.

\begin{figure}[t]
\centering
\includegraphics[width=\textwidth]{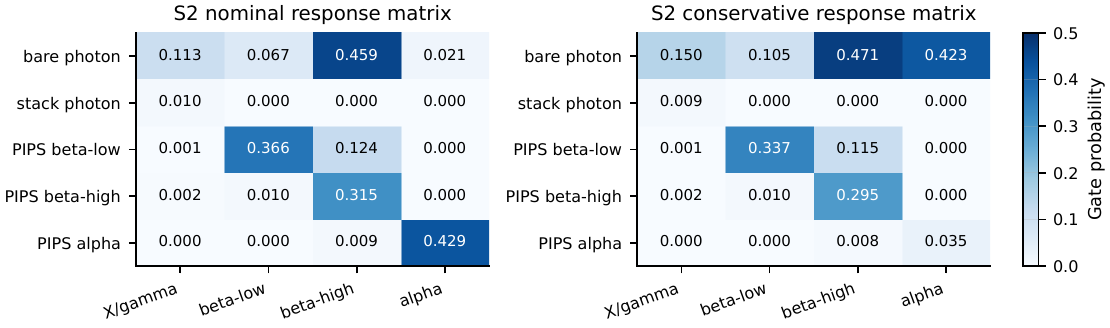}
\caption{Mean \Stwo\ response matrices after nominal and conservative electronics propagation. Columns are source classes and rows are exclusive measured gates. Values include live-time weighting.}
\label{fig:matrices}
\end{figure}

\Cref{fig:matrices} illustrates the nominal and conservative \Stwo\ response matrices. The rows are not pure labels. For example, high-energy beta contributes strongly to both the bare-photon and PIPS beta gates because primaries are distributed over the full incident face. Recovery uses the joint matrix rather than a single threshold decision.

\subsection{Level-3 same-model mixture closure}

\begin{figure}[t]
\centering
\includegraphics[width=\textwidth]{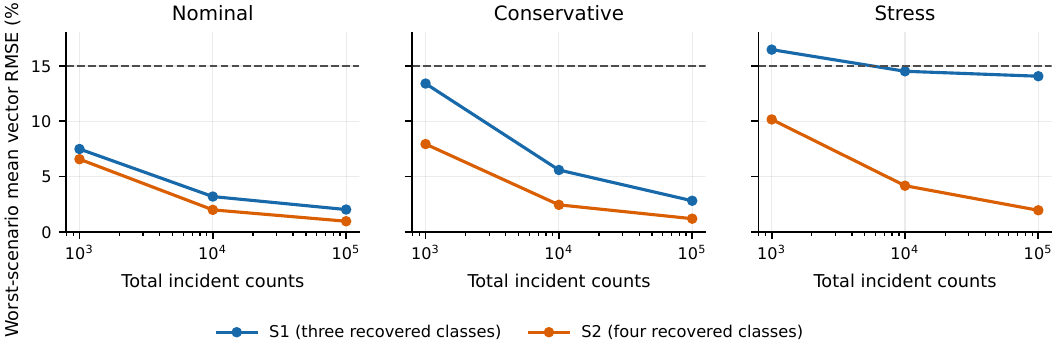}
\caption{Finite-count same-model closure. Each point is the largest mean vector RMSE among five labeled mixture scenarios; the dashed guide marks 15\%. \Sone\ estimates three classes after combining the low- and high-beta truth and therefore contains four distinct truth vectors; \Stwo\ estimates four classes.}
\label{fig:recovery}
\end{figure}

\Cref{fig:recovery,tab:level3} report the largest scenario-mean vector RMSE at each count level. Nominal \Stwo\ stays below the 15\% closure gate at all three levels and gives lower RMSE than the three-class \Sone\ while estimating one additional component. Under stress, \Sone\ reaches 16.48\% at \(10^3\) counts and remains near 14\% at \(10^5\), consistent with severe ill-conditioning rather than universal rank loss. \Stwo\ continues to improve with count, which is consistent with useful separation supplied by the backing-CZT row in the modeled response.

\begin{table}[t]
\centering
\caption{Worst-scenario mean vector RMSE as a percentage of total incident counts.}
\label{tab:level3}
\small
\begin{tabular}{llccc}
\toprule
Structure & Profile & \(N=10^3\) & \(N=10^4\) & \(N=10^5\) \\
\midrule
\Sone\ (three classes) & nominal & 7.49\% & 3.18\% & 2.01\% \\
\Sone\ (three classes) & conservative & 13.41\% & 5.59\% & 2.80\% \\
\Sone\ (three classes) & stress & 16.48\% & 14.52\% & 14.07\% \\
\Stwo\ (four classes) & nominal & 6.56\% & 1.98\% & 0.95\% \\
\Stwo\ (four classes) & conservative & 7.93\% & 2.43\% & 1.18\% \\
\Stwo\ (four classes) & stress & 10.16\% & 4.16\% & 1.94\% \\
\bottomrule
\end{tabular}
\end{table}

The worst nominal \Stwo\ case at \(N=10^3\) is the alpha-trace mixture. Its mean vector RMSE is 6.56\%, the 95th percentile is 14.23\%, and the 95th percentile of the largest component-wise error is 27.72 percentage points. Low-count performance is therefore distributional rather than captured by the mean alone. At \(N=10^5\), the largest scenario-mean RMSE is 0.95\%.

\subsection{Depth-relaxed U95 results}

Increasing the bare-CZT thickness changes both the analytical count chain and the independent transport response. At 662 keV, the mean probability of any bare-CZT deposition rises from 20.59\% in D10 to 25.63\% in D11.577 and 28.56\% in D12.570. The corresponding full-energy deposition probabilities are 4.92\%, 6.69\%, and 7.84\%. At the 3 MeV photon endpoint, the threshold-response probability increases from 6.61\% to 8.38\% and 8.88\%. These trends show that the added depth contributes usable photon interactions rather than only inactive package volume.

\Cref{tab:extensionresults} reports the U95 project metrics. D11.577 reaches 30.001 \(\sensunit\) in the raw analytical proxy but falls to 29.822 after the 0.9940 live fraction is applied. D12.570 retains 32.804 \(\sensunit\) after the same correction. Its independent-sum FWHM is 2.303\% at the 95th percentile. The propagated beta and alpha absolute-efficiency lower bounds are 35.679\% and 39.832\%; these values do not change with bare-CZT depth because the PIPS aperture and source geometry are fixed. All six range endpoints have nonzero response in every transport seed. D12.570 is therefore the only evaluated candidate that passes the complete project-performance gate under the application U95 model.

\begin{table}[t]
\centering
\caption{Project-performance metrics for the depth-relaxed \Stwo\ candidates under the U95 application profile. Sensitivity includes live-time correction; FWHM uses the independent sum channel.}
\label{tab:extensionresults}
\small
\begin{tabularx}{\textwidth}{YCCCCC}
\toprule
Candidate & \(S_{662}\) (\(\sensunit\)) & FWHM p95 & \(\eta_\beta\) lower & \(\eta_\alpha\) lower & Project gate \\
\midrule
D10 baseline & 24.917 & 2.344\% & 35.679\% & 39.832\% & Fail \\
D11.577 formal & 29.822 & 2.311\% & 35.679\% & 39.832\% & Fail \\
D12.570 margin & 32.804 & 2.303\% & 35.679\% & 39.832\% & Pass \\
\bottomrule
\end{tabularx}
\end{table}

The readout split is material to the resolution result. Direct digital summation of the independent layer channels gives 2.723\% for D11.577 and 2.819\% for D12.570, both above the project limit. The low-noise sum channel stays below 2.5\% for all three depths. Thus the passing design requires a dedicated sum path for spectroscopy; increasing the number of segmented channels alone is insufficient.

Layer-resolved continuous-energy classification improves the numerical separation of the four response columns. For D12.570, the test-matrix condition-number p95 falls from 27.10 with fixed gates to 5.466 with layered gradient boosting, and the mean test macro-F1 is 0.868. Across the registered \(10^4\)- and \(10^5\)-count mixture cells, the worst vector-RMSE p95 is 3.031\%, the worst component-error p95 is 5.970 percentage points, and alpha-trace misses and no-alpha false positives are both zero. The worst absolute total-bias p95 is 5.856\%, slightly above the internal 5\% diagnostic. This residual is relevant to calibration transfer, but it does not change the project-performance verdict because the task-book criteria do not specify a mixture-recovery tolerance.

\begin{figure}[t]
\centering
\includegraphics[width=\textwidth]{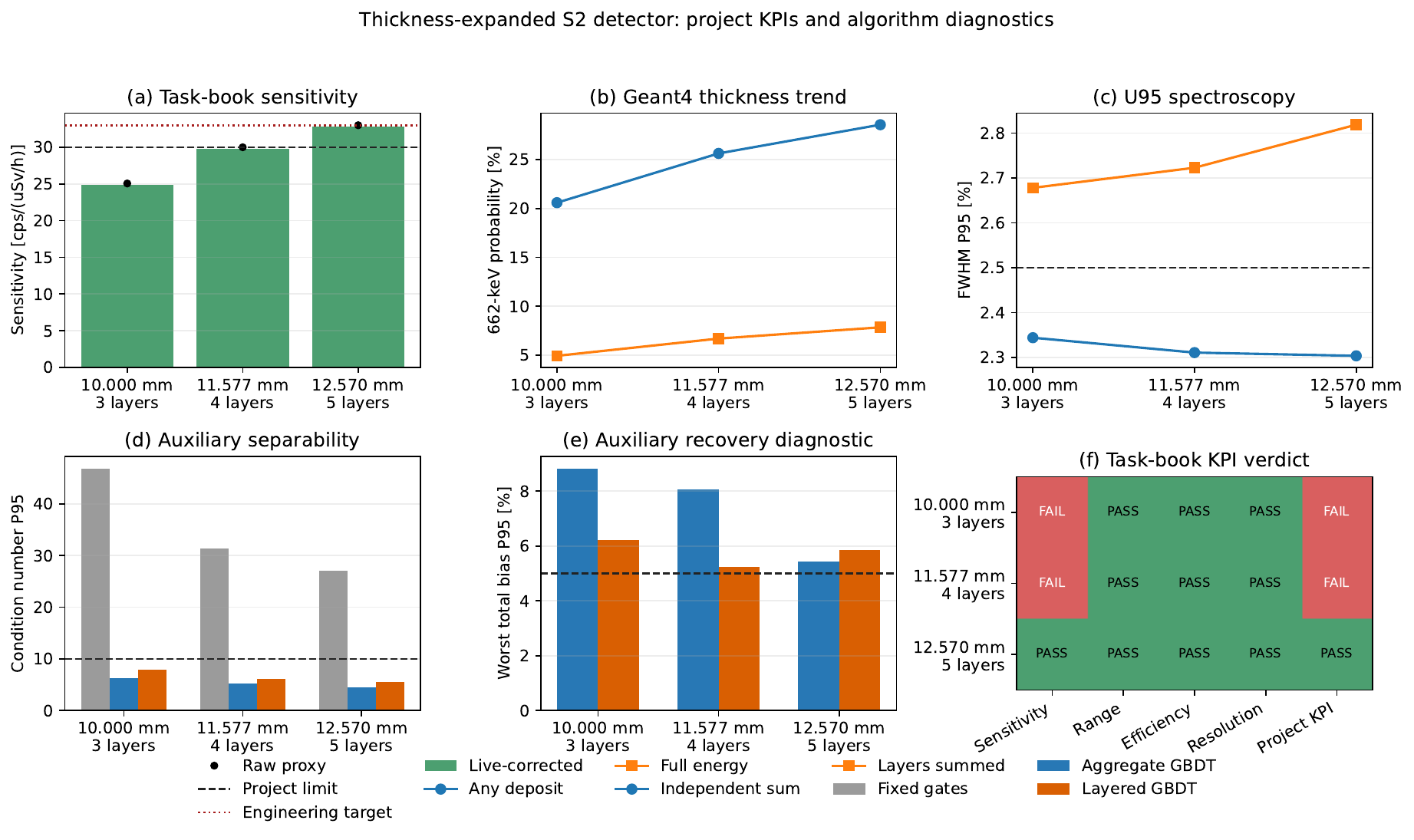}
\caption{Depth relaxation and U95 validation. Panels (a)--(c) show live-time-corrected sensitivity, fresh 662 keV transport trends, and the sum-channel/layer-sum resolution comparison. Panels (d) and (e) report auxiliary response-matrix and recovery diagnostics. Panel (f) applies only the original project detector-performance criteria; D12.570 is the passing candidate.}
\label{fig:thicknessu95}
\end{figure}

\section{Discussion}

\subsection{Which structure is preferable?}

The preferred structure follows from the required output. For X/gamma, aggregate beta, and alpha rates, \Sone\ is the lower-complexity baseline: it uses fewer channels and avoids a coincidence-dependent beta split. Its stress behavior, including three rank-deficient matrices among 30 draws, makes analog separation between the PIPS and bare-CZT paths the decisive implementation issue.

When low- and high-energy beta rates must be estimated separately, \Stwo\ is the stronger architecture. The backing CZT creates a large transport contrast between the sampled beta spectra and supplies a fifth observation for four unknown rates. The depth-relaxation study therefore extends \Stwo\ rather than reopening the architectural search. The instrument output is best defined as four calibrated mixed-field count-rate estimates; event-by-event labels are neither required nor implied. X rays and gamma rays remain one class unless spectral priors or additional observables are added.

\subsection{Metric-by-metric feasibility in the specified envelope}

The common V4 geometry fits the detector-head envelope: the active face occupies 38 by 18 mm, the occupied depth is 8.95 mm, and 1.05 mm remains along the depth axis. The model therefore supports geometric feasibility for the sensor stack itself. Detailed insulation, tolerances, flex routing, and thermal interfaces must fit within that reserve; the external high-voltage supply and acquisition electronics are outside the head definition.

Performance feasibility is narrower. Inside the 10 mm package, nominal \(H^*(10)\)-based photon sensitivity is 34.38--34.41 \(\sensunit\), but its robust counterpart is 25.04--25.07 \(\sensunit\). Every conservative electronics draw in the first campaign also exceeds 2.5\% FWHM. These results explain why a geometrically feasible screening point does not constitute a robust application design.

Relaxing only depth resolves this bottleneck in the present model. D12.570 increases external volume from 8.000 to 10.056 \(\mathrm{cm^3}\), a 25.7\% increase, while preserving the 40 by 20 mm face and charged-particle geometry. The added 2.570 mm supports two additional independently biased CZT layers. With the U95 electronics profile and the dedicated sum channel, sensitivity, resolution, charged-particle efficiency, and energy-range endpoints pass simultaneously. D11.577 is smaller but remains 0.178 \(\sensunit\) below the sensitivity threshold after live-time correction. The margin design is therefore preferred over the first raw crossing.

The two-stage result also clarifies the role of the algorithm. Gradient boosting does not create photon counts or improve the intrinsic FWHM. Thickness and layer bias supply the photon-interaction and charge-collection margin; the low-noise sum channel supplies the spectroscopic observable. The classifier uses the additional layer information to improve response-matrix conditioning and four-source rate estimation. Geometry, electronics, and inference address different constraints and cannot substitute for one another.

\subsection{Model scope and transfer conditions}

The calculations are most informative for architecture selection and specification flow-down. The sensitivity model combines traceable interaction data with a semi-empirical finite-aperture factor and assumed charge acceptance. The planar Hecht treatment resolves thickness, bias, and mobility--lifetime trade-offs, while an actual electrode weighting potential and crystal-defect map will determine the correction residual. Level 1 represents normally incident particles on ideal active faces; source capsule, self-absorption, air path, oblique incidence, package scatter, and environmental background enter the prototype response matrix. Level 2 spans designed electronics envelopes rather than a measured component population. Level 3 then closes the inverse problem with calibration and observations drawn from the same mean matrix, so its RMSE quantifies finite-count inversion within the model, not response transfer across model mismatch or drift.

These transfer conditions suggest a compact experimental sequence. Pulser and dark runs first establish leakage current, capacitance, ENC, gain, thresholds, timing jitter, dead time, and cross-channel coupling. A traceable \(^{137}\)Cs field then establishes the selected operational quantity and tests instrument response at the S-Cs photon energy of 662 keV \citep{iaea2000calibration,iso4037_3_2019}. \(^{241}\)Am and \(^{90}\)Sr--\(^{90}\)Y measurements establish the specified charged-particle efficiencies. Single-source measurements at several positions and angles populate each response-matrix column. Held-out two- and four-component mixtures, generated independently from the calibration set, test recovery at \(10^3\), \(10^4\), and \(10^5\) counts. Project acceptance should use the original sensitivity, resolution, range, and efficiency criteria. Conditioning and mixture-recovery limits can be registered separately as algorithm qualifications.




\section{Conclusion}

The fixed-budget study identifies two physically feasible architectures within \(40\times20\times10\ \mathrm{mm^3}\). \Stwo\ is preferred when low- and high-beta rates must be estimated separately because its backing-CZT trigger probability changes from 2.12\% to 70.47\% across the sampled beta spectra. The same study also exposes the limiting constraints: the robust \(H^*(10)\) sensitivity is only 25.04--25.07 \(\sensunit\), and the conservative electronics profile does not meet 2.5\% FWHM.

The depth-relaxed study keeps the \Stwo\ face allocation and charged-particle branch, increases package depth to 12.570 mm, and divides 7.870 mm of bare CZT into five independently biased layers. A fresh 2.40-million-history, five-seed transport campaign and the application U95 electronics model give a live-time-corrected sensitivity of 32.804 \(\sensunit\), a sum-channel FWHM p95 of 2.303\%, beta/alpha absolute-efficiency lower bounds of 35.679\%/39.832\%, and nonzero response throughout the specified energy endpoints. D12.570 therefore passes the original detector-performance criteria in the U95 model. Layer-resolved gradient boosting reduces the response-matrix condition-number p95 to 5.466 and supports four-rate recovery, while its 5.856\% worst total-bias p95 remains an auxiliary calibration diagnostic. 


\section*{Acknowledgments}
{\sloppy
This work was supported by the National Key Research and Development Program of China (Grant No.~2024YFB3213203).\par}



\clearpage
\appendix
\section{Frozen electronics profiles}

\begin{center}
\captionof{table}{Inputs to the Level-2 electronics envelopes.}
\small
\begin{tabular}{lccc}
\toprule
Parameter & Nominal & Conservative & Stress \\
\midrule
Si ENC (keV) & 1.0 & 2.0 & 4.0 \\
Bare-CZT ENC (keV) & 2.5 & 5.0 & 10.0 \\
Back-CZT ENC (keV) & 3.0 & 6.0 & 12.0 \\
Gain dispersion (relative) & 0.0025 & 0.0050 & 0.0100 \\
CCE dispersion (relative) & 0.0045 & 0.0085 & 0.0125 \\
Threshold drift sigma (keV) & 1.5 & 3.0 & 6.0 \\
Crosstalk fraction & 0.002 & 0.007 & 0.015 \\
Dead time (\(\mu\)s) & 2 & 5 & 10 \\
Event rate (Hz) & 1,000 & 5,000 & 10,000 \\
Timing jitter (ns) & 50 & 100 & 150 \\
\bottomrule
\end{tabular}
\end{center}

\section{Computational checks and requirement reconciliation}

\begin{center}
\captionof{table}{Computational checks and requirement reconciliation across the fixed-budget and depth-relaxed stages.}
\small
\begin{tabularx}{\textwidth}{Yp{32mm}p{32mm}}
\toprule
Gate & Threshold & Result \\
\midrule
Minimum detected events in each key Level-1 metric & \(\geq25\) & minimum observed 254; pass \\
Nominal 662 keV FWHM & \(\leq2.5\%\) for both & pass \\
Nominal mean response-matrix condition number & \(\leq10\) for both & 5.69 / 9.50; pass \\
Conservative live fraction & \(\geq0.95\) for both & pass \\
Nominal \Stwo\ same-model closure RMSE & \(\leq15\%\) scenario mean & 6.56\%; pass \\
Fixed-budget robust \(H^*(10)\) sensitivity & \(\geq30\ \sensunit\) & 25.04--25.07; not met \\
Fixed-budget conservative 662 keV FWHM & \(\leq2.5\%\) for both & 3.04 / 2.98\%; not met \\
Depth-relaxed transport campaign & all jobs and energy conservation & 210/210 jobs; 2.40 million events; pass \\
D12.570 U95 operational sensitivity & \(\geq30\ \sensunit\) & 32.804; pass \\
D12.570 U95 sum-channel FWHM p95 & \(\leq2.5\%\) & 2.303\%; pass \\
D12.570 beta/alpha efficiency lower bounds & \(\geq35\%/25\%\) & 35.679\%/39.832\%; pass \\
D12.570 range endpoints & 20 keV--3 MeV; 3--7 MeV; 155 keV--3.5 MeV & nonzero in all seeds; pass \\
D12.570 project-performance gate & all original detector criteria & pass \\
D12.570 auxiliary total-bias p95 & internal 5\% diagnostic & 5.856\%; above diagnostic \\
\bottomrule
\end{tabularx}
\end{center}

\clearpage
\bibliographystyle{unsrtnat}
\bibliography{references}

\end{document}